\documentstyle[11pt,mypasp,twoside,epsf]{article} 
\def\edcomment#1{\iffalse\marginpar{\raggedright\sl#1\/}\else\relax\fi} 
\reversemarginpar 

\begin{document} 


\title{Correlations of Globular Cluster Properties: \\
       Their Interpretations and Uses}

\author{S. G. Djorgovski} 
\affil{California Institute of Technology, Pasadena, CA 91125, USA} 
\author{P. C\^ot\'e} 
\affil{Dept.~Phys.~\& Astronomy, Rutgers U., New Brunswick, NJ 08854, USA} 
\author{G. Meylan} 
\affil{Space Telescope Science Institute, Baltimore, MD 21218, USA} 
\author{S. Castro} 
\affil{IPAC/SSC, Caltech, Pasadena, CA 91125, USA} 
\author{L. Federici, G. Parmeggiani, C. Cacciari, F. Fusi Pecci} 
\affil{INAF -- Osservatorio Astronomico di Bologna, Bologna, I-40127, Italy} 
\author{R.M. Rich} 
\affil{Dept.~of Physics \& Astronomy, UCLA, Los Angeles, CA 90095, USA} 
\author{T. Jou}
\affil{California Institute of Technology, Pasadena, CA 91125, USA}

\begin{abstract} 
Correlations among the independently measured physical properties of globular
clusters (GCs) can provide powerful tests for theoretical models and new
insights into their dynamics, formation, and evolution.  We review briefly
some of the previous work, and present preliminary results from a comparative
study of GC correlations in the Local Group galaxies.  The results so far
indicate that these diverse GC systems follow the same fundamental correlations,
suggesting a commonality of formative and evolutionary processes which produce
them.
\end{abstract}

\section{Introduction} 

Understanding of the physics, formation, and evolution of any type of
astronomical objects or systems, including globular clusters (GCs) as a family,
must rest on a solid, quantitative empirical foundation.  In the order of
an increasing information content, the first
step is a definition of characteristic values, e.g., a
typical mass or luminosity or half-light radius, etc.  The next step is a
determination of distribution functions for various physical quantities,
e.g., the luminosity function.  Finally, most information can be obtained 
from non-trivial correlations of independently measured or derived quantities.
Such correlations are products of some physical or evolutionary processes,
and as such contain valuable clues towards the understanding of the objects
in question.

In the context of globular clusters, previous studies include, e.g.,
Chernoff \& Djorgovski (1989), 
Djorgovski (1991, 1995, 1996),
Djorgovski \& Meylan (1994),
Bellazzini et al. (1996),
Bellazzini (1998),
McLaughlin (2000),
and numerous papers by Sidney van den Bergh, e.g, van den Bergh (1996).
For a comparison of GCs with other old stellar systems, see, e.g.,
Kormendy (1985) or Djorgovski (1993).
On-line compilations of Galactic GC data useful for such studies include
Harris (1996),
\footnote{http://physun.physics.mcmaster.ca/Globular.html}
which is an updated superset of the data
\footnote{http://www.astro.caltech.edu/$\sim$george/glob/data.html}
presented in the First Ivan King Festschrift (eds. Djorgovski \& Meylan 1993).
Both are now well overdue for a major revision and updates, including
Hipparcos-based distances, uniform IR photometry and reddenings derived
from the 2MASS data, better core parameters from the HST-based surface
photometry, etc.

Additional insights can be gained by comparing correlations of GC properties
in GC systems of different galaxies.  Currently, we are in practice limited
to the galaxies of the Local Group.  Previous studies include, e.g.,
Fusi Pecci et al. (1994),
Djorgovski et al. (1997),
Dubath \& Grillmair (1997),
Dubath et al. (1997),
Meylan et al. (2001),
Barmby et al. (2002),
etc.
In this paper we present preliminary results from a new study of dynamical
correlations for GC systems in several Local Group galaxies.

\section{Correlations for Galactic GCs: A Brief Overview} 

A generic expectation from the differences in corresponding dynamical time
scales is that the evolution at the core radius scales would be much faster
than at the half-light radius scales.  The former then reflects mainly the
evolution towards the core collapse (and any self-similar behavior before
and after the core collapse), whereas the latter reflects more the initial
conditions and a long-term evolution.  Position of a GC in the Galactic
potential modulates the evolutionary effect through the effect of dynamical
shocks due to the disk and bulge passages, in the sense that clusters
exposed to more frequent and stronger shocks evolve faster.  This introduces
a secondary dependence on the distance to the Galactic center and plane in
many of the observed trends and correlations.  All this was well documented
in the references cited above, and at least the qualitative agreement
between the observations and theory is striking.

\begin{figure}
\null
\plotone{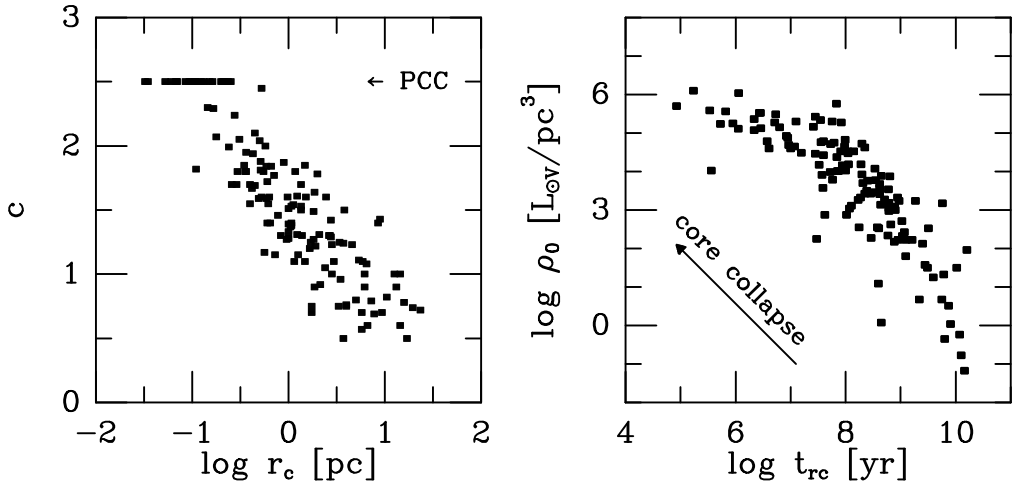}
\caption{
Examples of core parameter correlations, driven by the evolution towards the
core collapse.  $Left:$  Correlation between two observables, the King 
concentration parameter $c$, and the core radius $r_c$.  Clusters with an
unresolved (from the ground) post-core-collapse (PCC) morphology have been
assigned $c = 2.5$.  $Right:$  An illustrative correlation between two
derived quantities, the central luminosity density, $\rho_0$, and the
central relaxation time, $t_{rc}$.
}
\end{figure}

In a simple core collapse picture, as
$t \rightarrow t_c$ (time of the maximum collapse),
the core radius $r_c \rightarrow 0$,
the concentration parameter $c \rightarrow \infty$,
the central density $\rho_0 \rightarrow \infty$, and
the central surface brightness $I_0 \rightarrow \infty$,
in a self-similar manner;
see, e.g., Meylan \& Heggie (1997) for a review and references.
The observed cluster properties at the core scale are thus driven towards
a (nearly) 1-parameter sequence corresponding to the relative dynamical
time away from the core collapse (Fig. 1).  This is indeed consistent with the
observations: the predicted trend in simple single-component, Fokker-Planck
core collapse models is 
$\rho_0 \sim r_c ^{~ -2.23}$,
whereas the observed trends are
$I_0 \sim r_c ^{~ -1.8 \pm 0.2}$ and
$\rho_{0,lum} \sim r_c ^{~ -2.6 \pm 0.2}$.
This implies for the core mass
$m_{core} \sim r_c ^{~ 0.3 \pm 0.2} (M/L)$, 
i.e., a nearly constant core mass, possibly slightly diminishing due to
the evaporation of high-energy stars, and/or becoming slightly darker due
to the mass segregation of heavy stellar remnants.

No such trends are seen at the half-light radius scale, where the
dynamical range of the relaxation time scales relative to the Galactic age
are is much smaller ($t_{rh} \sim 10^8 - 10^{10}$ yr, whereas for the core 
relaxation time scales
$t_{rc} \sim 10^5 - 10^{10}$ yr), so that the internal spread of GC
properties dominates over the dynamical evolution effects.

Likewise, there are only a few noisy trends with the cluster luminosity
($\sim$ mass), which are apparent only when the data are binned: more
luminous clusters tend to be more concentrated,
$r_c \sim L ^{-0.5 \pm 0.25}$,
$\rho_0 \sim L ^{2 \pm 1}$.

One can also estimate the rough tidal radii $r_t$ from the observed surface
brightness profiles, and the mean cluster densities $\rho_t$ within the $r_t$.
One finds a mean trend with the present Galactocentric radius $R_{GC}$ of
$r_t \sim R_{GC} ^{~ 0.37 \pm 0.05}$,
as intuitively expected, and also
$\rho_t \sim R_{GC} ^{~ -1.6 \pm 0.2}$,
close to the mean density law for the dark halo.  Using a simple theory
(Innanen et al. 1983) and an assumed Galactic rotation curve, it is then
possible to estimate the perigalactic radii, $R_{peri}$ (see Djorgovski 1996
for more details).  Intriguingly, the ratio $R_{GC}/R_{peri}$ peaks near the
unity, with a long tail, suggesting that most GCs today are on nearly circular
orbits.  This could be a survival selection effect, or a reflection of the
initial conditions, or a combination of the two.

\begin{figure}
\null
\plotone{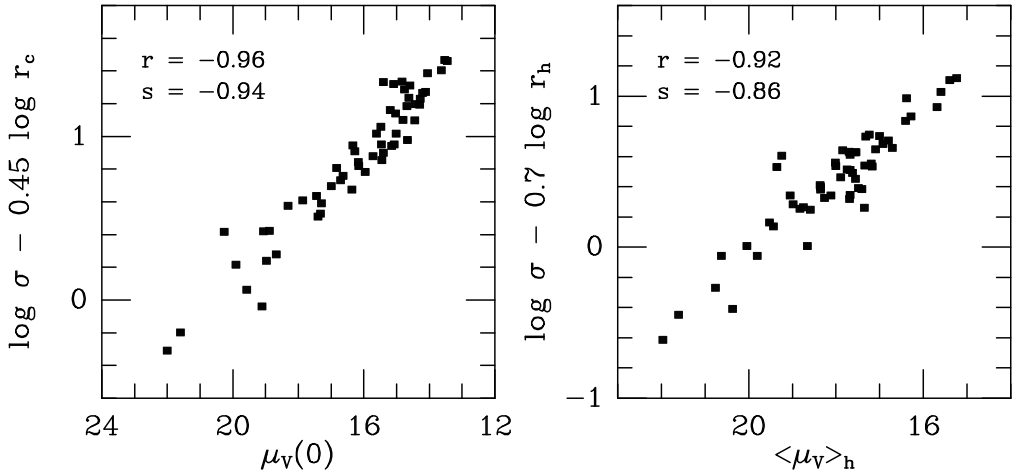}
\caption{
The Fundamental Plane (FP) bivariate correlations for GCs: an optimal
combination of the central velocity dispersion ($\sigma$) and core radius
($r_c$) is correlated with the central surface brightness in the $V$ band
(left panel), and a different combination of $\sigma$ and the half-light
radius $r_h$ with the mean surface brightness within $r_h$ (right panel).
Pearson (r) and Spearman (s) correlation coefficients are indicated in
each panel.  The residual scatter is completely accounted by the measurement
errors.
}
\end{figure}

Possibly the most interesting monovariate correlations are those including
core velocity dispersions, $\sigma$:
$\sigma \sim L ^{0.6 \pm 0.15}$ (or equivalently, $L \sim \sigma ^{5/3}$),
$\sigma \sim I_0 ^{~ 0.5 \pm 0.1}$, and
$\sigma \sim I_h ^{~ 0.45 \pm 0.05}$, where
$I_h$ is the mean surface brightness within the half-light radius
(see Figs. 3--4).
The origin of these correlations is still not understood, but several
possibilities exist; see, e.g., Djorgovski \& Meylan (1994) for a
discussion.  These correlations may be reflecting the formative mechanisms
of GCs, and they should be reproduced by any successful model of GC formation
and long-term evolution.  They are also very different from the corresponding
correlations for elliptical and dwarf (dE, DSph) galaxies (Djorgovski 1993).

\begin{figure}
\null
\plotone{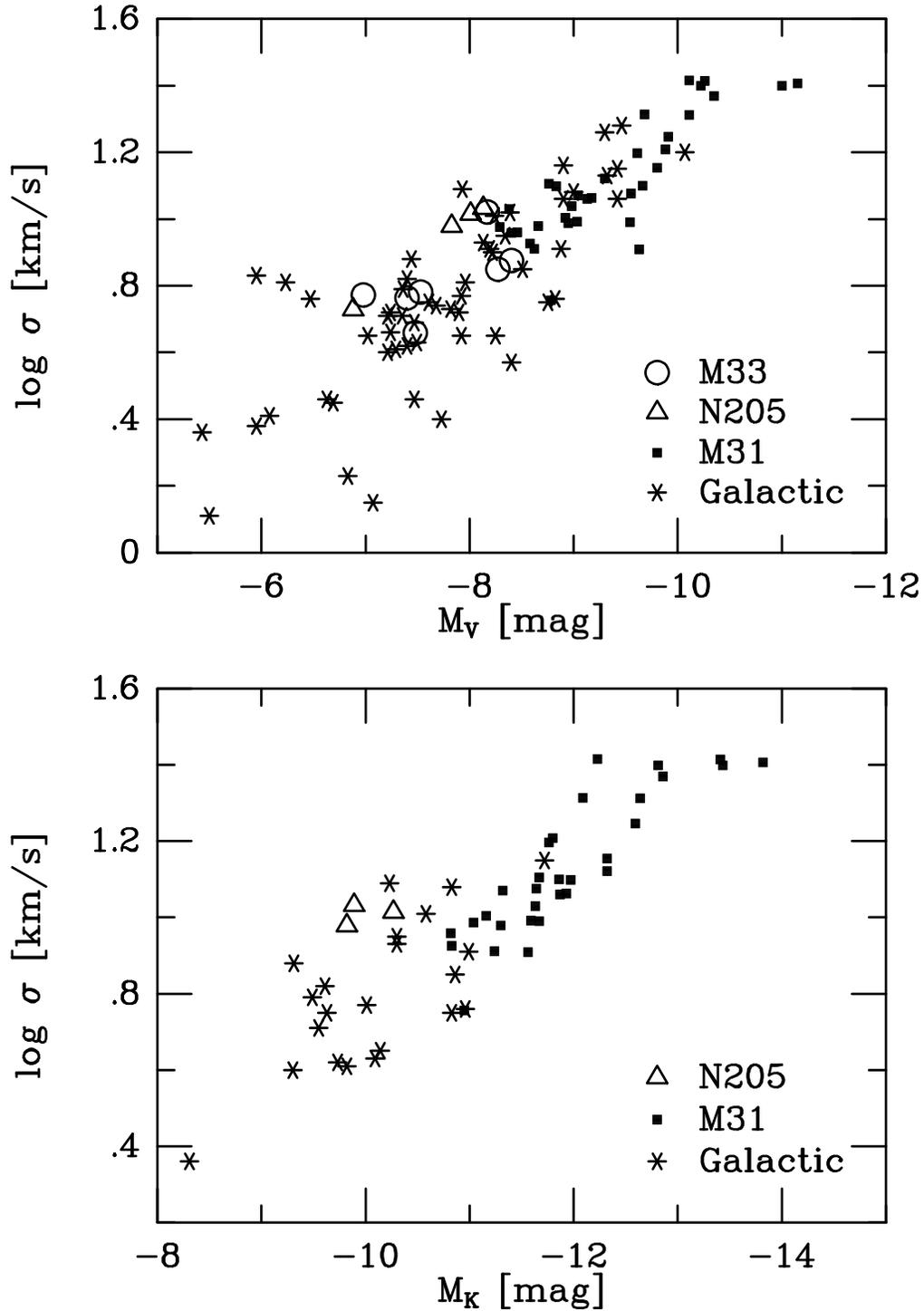}
\caption{
Luminosity -- velocity dispersion correlations in the $V$ (top) and $K$
(bottom) bands for GCs in 4 Local Group galaxies, encoded with different
symbols as indicated in the figure.  The clusters in N205 appear to be
slightly underluminous for their velocity dispersions, but this may be
due to a systematic error.  Otherwise, all of these GC systems appear to
follow the same dynamical correlations.
}
\end{figure}

It is notable that GC metallicities do not seem to correlate with anything,
in contrast to other old stellar systems.  This strongly suggests that GCs
were not self-enriched.

Another interesting question is how many independent physical parameters
control the observable properties of GCs?  Following the pioneering study 
by Brosche \& Lentes (1984), more modern data sets suggest that there are
at most 6 significant parameters among 9 or 10 independently measured
observables (Djorgovski 1991, Djorgovski \& Meylan 1994).  If one considers
just the observed photometric, structural, and dynamical parameters at both
core and half-light scales, the statistical dimensionality of the data is 3,
or 4 if the $(M/L)$ ratios are included.  This is exactly as expected for a
manifold of King (1966) models.

\begin{figure}
\null
\plotone{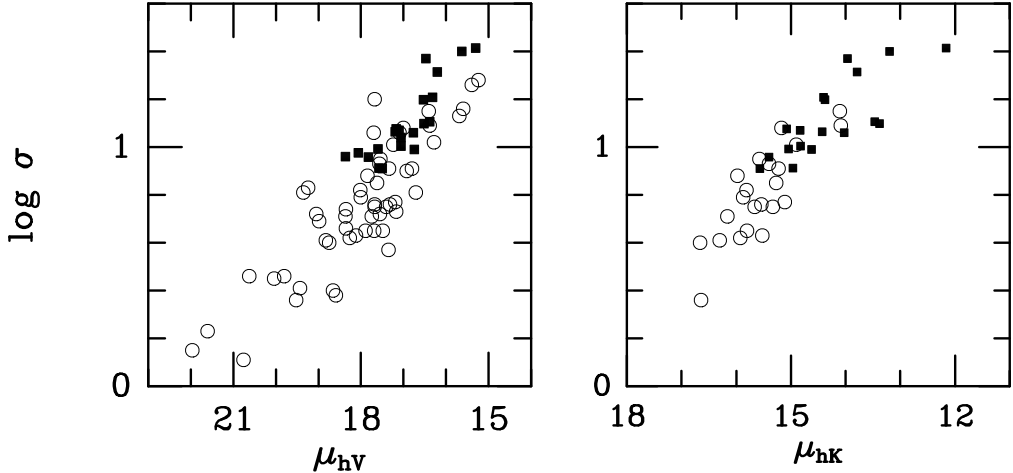}
\caption{
Correlations between the velocity dispersion $\sigma$ and the mean surface
brightness within the half-light radius in the $V$ band ($\mu_{hV}$, left
panel) and in the $K$ band ($\mu_{hK}$, right panel) for the Galactic GCs
(open circles) and M31 GCs (solid squares).  The M31 extends to higher
luminosities and masses, and is selection-limited at the low end, but in
the overlap region the two GC systems follow the same correlations.
}
\end{figure}

Constraining the input parameter set to either core or half-light ones, brings
the statistical dimension of the data set to 2, i.e., the triplets of
observables are connected by bivariate correlations (Djorgovski 1995; see
Fig. 2).  This
is the equivalent of the Fundamental Plane (FP) of elliptical galaxies.
For the core parameters, the scaling relation is:
$r_c \sim \sigma ^{2.0 \pm 0.2} I_0 ^{~ -1.1 \pm 0.1}$.
If we assume a structural homology (probably a good assumption for the GC
cores), the virial theorem implies:
$r \sim \sigma ^2 I ^{-1} (M/L) ^{-1}$.
Thus, the FP of GC cores implies that their $(M/L)$ ratios are (nearly?)
constant; this is the $sharpest$ observational constraint on the constancy
of GC $(M/L)$ ratios to date.
In contrast, for the half-light parameters, the scaling relation is:
$r_h \sim \sigma ^{1.45 \pm 0.2} I_h ^{~ -0.85 \pm 0.1}$.
This is remarkably similar to the FP of E-galaxies, and is almost certainly
a consequence of the non-homology of GC structures.

An alternative look at the FP of GCs was provided by McLaughlin (2000).
While recognizing the equivalence of his FP with the one described above,
his preferred scaling relation is the expression of the binding energy,
$E_b \sim L ^{2.05} R_{GC} ^{~ -0.4}$, where the 2nd term effectively corrects
for the known dependences of the GC parameters on their position in the
Galaxy.  However, we note that the binding energy can be written as:
$E_b = L ^2 (M/L)^2 r_h ^{~ -1} f(c)$, where $f(c)$ is a slow function 
of the cluster concentration.  If the $(M/L)$ ratios are indeed constant,
and knowing that $r_h$ does not correlate much with anything, it is then
not surprising that the observed $R_{GC}$-corrected scaling is
$E_b \sim L ^{2.05}$.

\begin{figure}
\null
\plotone{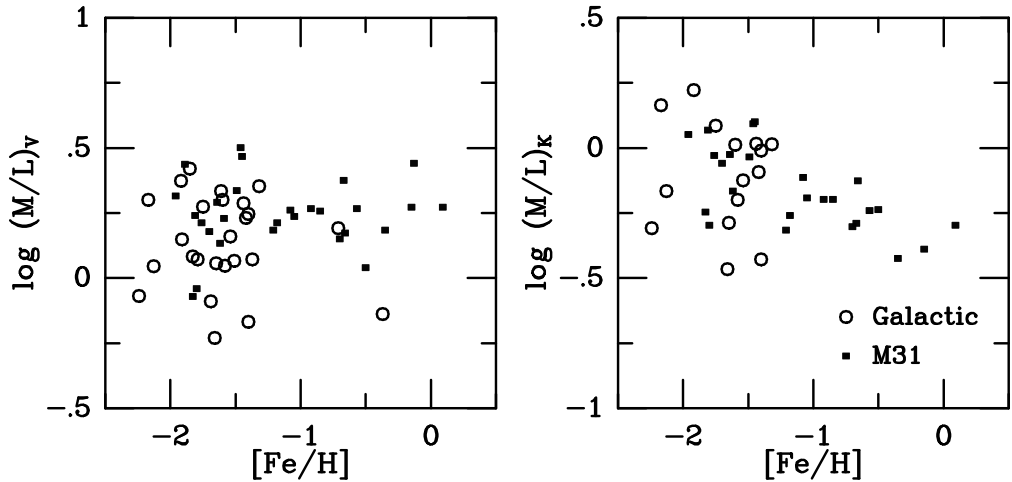}
\caption{
Dependence of the derived $(M/L)$ ratios (in Solar $V$ band units) on the
metallicity, for the Galactic GCs (open circles) and M31 GCs (solid squares),
in the $V$ (left) and $K$ (right) bands.  The larger observed scatter for the
Galactic GCs may be caused in part by the distance errors, whereas all M31 GCs
are effectively at the same distance.  While no trend is seen in the $V$ band
(as implied by the FP; see above), there is a trend in the $K$ band in the
sense of more metal-rich clusters having more luminous stellar populations
at a given mass.
}
\end{figure}

\section
{A Comparison of Dynamical Correlations for GCs in Five Local Group Galaxies:
Some Preliminary Results} 

Correlations between the velocity dispersion and other parameters ($L$, $I_0$,
$I_h$, and the FP) may probe directly the physics and formative processes of
GCs, their homogeneity (or lack thereof) in different galaxies, and can be also
used as new distance indicator relations for their parent galaxies, providing
an independent check of other distance scales.  
Fusi Pecci et al. (1994) and Barmby et al. (2002) established that the
structural and photometric parameters of GCs in M31 as observed with the HST
occupy the same portion of the parameter space as the Milky Way GCs.
Following on the initial studies by Dubath \& Grillmair (1997) and
Djorgovski et al. (1997), we set to explore in more detail the dynamical
correlations for GCs in 5 Local Group galaxies: the Milky Way, M31, M33,
N185, and N205.

The extra-Galactic sample consists of HST imaging of GCs in M31, M33, N185,
and N205 available as of the early 2002, for which we have done surface
photometry (Federici, Parmeggiani, et al., in prep.).  Velocity dispersions
have been measured using Echelle spectra obtained at the Keck-I telescope with
the HIRES instrument, as described, e.g., in Djorgovski et al. (1997); these
measurements will be presented in detail elsewhere (C\^ot\'e et al., in prep.).

Figures 3--5 illustrate some of our preliminary results.  Our first conclusion
is that GCs in these different galaxies follow essentially the same dynamical
correlations.  This suggests a common set of physical mechanisms affecting
their formation and evolution, despite a broad range of their host galaxy
properties.  The observed dependence of the $(M/L)$ ratios on the metallicity,
especially in the $K$-band (Fig. 5), represents a useful observational
constraint on the models of old stellar populations.
In a future paper (Djorgovski et al., in prep.) we will present a more complete
and detailed analysis, including the FP correlations for these GC systems.

Even with the resolution
of the HST, such studies cannot be pushed much beyond the Local Group.
For example, Harris et al. (2002) present an excellent study of GCs in
Cen A = N5128.

\acknowledgments

We wish to thank our collaborators, and the staff of W.M. Keck Observatory
for their expert help during our observing runs.  This work was supported by
grants from the NSF, NASA/STScI (GO-6671 and AR-8735), and private donors
in the U.S., and ASI and MURST in Italy.  PC was supported in part by a
Fairchild Fellowship, and TJ by a SURF Fellowship at Caltech.

\end{document}